\documentclass[a4paper,12pt]{article}
\usepackage[margin=2.0cm]{geometry}

\RequirePackage{amsmath}
\RequirePackage{braket}
\RequirePackage{amssymb}
\RequirePackage{amsfonts}
\RequirePackage{latexsym}
\RequirePackage{graphicx}
\RequirePackage{color}
\RequirePackage{indentfirst}
\RequirePackage[colorlinks,citecolor=blue,urlcolor=blue,linkcolor=blue]{hyperref}

\newcommand{\sTr}{\,\mbox{sTr}\,}

\newcommand{\al}{\alpha}
\newcommand{\be}{\beta}
\newcommand{\ga}{\gamma}
\newcommand{\Ga}{\Gamma}
\newcommand{\de}{\delta}

\newcommand{\vp}{\varepsilon}

\newcommand{\om}{\omega}

\newcommand{\pa}{\partial}

\newcommand{\beq}{\begin{eqnarray}}
\newcommand{\eeq}{\end{eqnarray}}
\newcommand{\eq}[1]{(\ref{#1})}
\newcommand{\n}[1]{\label{#1}}
\newcommand{\nn}{\nonumber}

\newcommand{\bAEA}{background average effective action }
\newcommand{\BFM}{background field method }

\begin{document}

\begin{center}

{\Large \bf
Gauge invariance of the background average effective action}

\vskip 6mm

\small{
\text{Peter M. Lavrov}$^{a,b,c}$,
%
\text{Eduardo Antonio dos Reis}$^a$,
%
\text{Tib\'{e}rio de Paula Netto}$^d$,
%
\text{Ilya L.Shapiro}$^{a,b,c}$
\footnote{ E-mails: lavrov@tspu.edu.ru (Peter M. Lavrov), eduardoreis@ice.ufjf.br (Eduardo Antonio dos Reis), tiberio@sustech.edu.cn (Tib\'{e}rio de Paula Netto), shapiro@fisica.ufjf.br (Ilya L.Shapiro).}

\vskip 4mm

(a){ Departamento de F\'{\i}sica,  ICE,  Universidade Federal de Juiz de Fora,\\
36036-330 Juiz de Fora, \ MG, \ Brazil}
\vskip 2mm

(b) {Department of Mathematical Analysis, Tomsk State Pedagogical \\
University,  634061, Tomsk, Russia}
\vskip 2mm

(c){ National Research Tomsk State University, 634050 Tomsk, Russia}
\vskip 2mm

(d) {Department of Physics, Southern University of Science and Technology,\\ 
Shenzhen 518055, China}

} 

\vskip 4mm

\begin{quotation}
\begin{abstract}
\noindent
Using the background field method for the functional renormalization group approach in the case of a generic gauge theory, we study the background field symmetry
and gauge dependence of the background average effective action, when the regulator action depends on external fields. The final result is that the symmetry of the average effective
action can be maintained for a wide class of regulator functions, but in all cases the dependence of the gauge fixing
remains on-shell. The Yang-Mills theory is considered as the main particular example.
\end{abstract}
\end{quotation}

\end{center}
\vskip 4mm

\section{Introduction}
\label{intro}

One of the most prospective non-perturbative approaches in
quantum field theory (QFT) is the functional (or exact)
renormalization group (FRG), which is based on the Wetterich
equation for the average effective action \cite{Wetterich,TimMorris}
(see the reviews
\cite{FRG-review,FRG-review-1,FRG-review-2,FRG-review-3}
and textbook \cite{Wipf} for an introduction to the subject).
The application of FRG to gauge theories was extensively
discussed, including in the recent work \cite{Wetterich-2018}.
The considerations in the last and many other papers are based
on the background field method, which enables one to maintain
the gauge invariance for the Yang-Mills (or gravitational) field
explicitly in the effective action. The background field method
is, in general, a useful formalism in the theory of gauge fields,
and that is why it attracted a very special attention recently,
see e.g. \cite{Barv,BLT-YM,FT,Lavrov}. The application of this
method to the average effective action has been done long ago
\cite{Reuter} (see also the recent work \cite{CODELLO}), but
in our opinion there are some important aspects of the problem
which should be explored in more details.

The main problem of FRG applied to the gauge theories is
that the dependence on the choice of the gauge fixing condition
does not disappear on-shell \cite{LSh}, as it is the case in
the usual perturbative QFT. As a result of the on-shell
gauge fixing dependence, the $S$-matrix of the theory is
not well defined, except at the fixed point, where the effective
average action coincides with the usual effective action. One can
expect that the renormalization group flow in the Yang-Mills theory
will also manifest a fundamental gauge dependence, and this
certainly shadows the interpretation of the results obtained
within the FRG approach in the gauge theories.

In order to better understand the situation with the gauge
symmetry at the quantum level and with the gauge dependence,
it is important to analyze the mentioned problems in the background
field method, that is the main purpose of the present communication.
In what follows we consider both gauge invariance and gauge fixing
dependence for the effective average action.

The paper is organized as follows. In Sec.~\ref{Sec2} we give a brief
description of the background field formalism in non-Abelian gauge
theories and the gauge independence of vacuum functional in this
method. In Sec.~\ref{Sec3} the background field symmetry is analyzed
within the  \BFM- based functional renormalization group approach.
The regulator functions are dependent on the external (background)
fields but are chosen not to be invariant under gauge transformations
of external vector field.  In Sec.~\ref{Sec4} we present a solution
to the background field symmetry of \bAEA with regulator functions
which are invariant under gauge transformations of the external
field. In Sec.~\ref{Sec5} the gauge dependence problem of the
\bAEA is considered.  Finally, we discuss the results and draw
our conclusions in Sec.~\ref{Sec6}.

Our notations system mainly follows the DeWitt's book
\cite{DEWITT}. Also, the Grassmann parity of a quantity $\,A\,$
is denoted $\,\vp(A)$.

\section{Background field formalism}
\label{Sec2}

We start by making a brief review of the background field formalism for
a gauge theory describing by an initial action $S_0(A)$ of fields $A=\{A^i\}$,
$\varepsilon(A^i)=\varepsilon_i$
 invariant under gauge transformations
\beq
\label{YM}
S_{0,i}(A)R^i_{\alpha}(A)=0, \quad \delta
A^i=R^i_{\alpha}(A)\xi^{\alpha},
\eeq
where $R^i_{\alpha}(A)$,
$\varepsilon(R^i_{\alpha}(A))=\varepsilon_i+\varepsilon_{\alpha}$
are the generators of gauge transformations, $\xi^{\alpha}$,
$\varepsilon(\xi^{\alpha})=\varepsilon_{\alpha}$ are
arbitrary functions. In general, a set of
fields $\,A^i=(A^{\alpha k},A^m)\,$ includes fields $A^{\al k}$
of the gauge sector and also fields $A^m$ of the matter sector
of a given theory. We
assume that the generators $ R^i_{\alpha}= R^i_{\alpha}(A)$
satisfy a closed algebra with structure coefficients
$F^\gamma_{\alpha\beta}$ that do not depend on the fields,
\begin{align}
\label{AlgebraAi}
R^i_{\al,j}R^j_{\beta} - (-1)^{\vp_\al \vp_\be} R^i_{\be ,j}R^j_{\al}  = 
 - R^i_\ga F^\ga_{\al\be},
\end{align}
where we denote the right functional derivative by $\de_r X / \de A^i = X_{,i}$. The structure coefficients
satisfy the symmetry properties
$\,F^\ga_{\al\be}=-(-1)^{\vp_\al \vp_\be}F^\ga_{\be\al}$.
We assume as well that the generators are linear
operators in $A^i$, $R^i_{\alpha}(A)=t^i_{\alpha j}A^j+r^i_{\alpha}$.

We apply the background field method (BFM) \cite{DWITT,FADEEV,ABBOTT}
replacing the field $\,A^i\,$ 
by $\,A^{i} + \mathcal{B}^{i}\,$ in the classical action
$S_0(A)$,
\beq
\n{BYM}
S_{0}(A) \,\,\longrightarrow \,\, S_{0}(A +\mathcal{B})\,.
\eeq
Here
$\,\mathcal{B}^{i}$ are  external (background) vector fields
being not equal to zero only in the gauge sector.
The action $\,S_{0}(A +\mathcal{B})\,$ obeys the gauge invariance
in the form
\beq
\n{BGI}
\de  S_{0}(A +\mathcal{B}) = 0,
\qquad
\de  A^{i} = R^i_{\alpha}(A+\mathcal{B})  \xi^{\alpha}\,.
\eeq

Through the Faddeev-Popov quantization \cite{FP} the field configuration
space is extended to
\beq
\n{Phi}
\phi^A = (A^i,B^{\alpha},C^{\alpha}, \bar{C}^{\alpha}),
\qquad
\vp(\phi^A) = \vp_A,
\eeq
where $C^{\alpha}$, $\bar{C}^{\alpha}$ are the Faddeev-Popov ghost and antighost
fields, respectively, and $B^a$ is the auxiliary (Nakanishi-Lautrup) field.
The Grassmann parities distribution are the following
\beq
\vp(C^{\alpha}) = \vp(\bar{C}^{\alpha}) =\vp_{\alpha}+ 1,
\qquad
\vp(B^{\alpha}) = \vp_{\alpha}\,.
\eeq
%
The corresponding Faddeev-Popov action $ S_{FP} (\phi, \mathcal{B})$
in the singular gauge fixing has the form~\cite{FP}
\beq
\n{FP}
S_{FP} (\phi, \mathcal{B})
 =
 S_{0}(A +\mathcal{B})
+
S_{gh} (\phi, \mathcal{B})
+
S_{gf} (\phi, \mathcal{B})
\eeq
where
\begin{align}
\n{Sgh}
S_{gh} (\phi, \mathcal{B})
&=
\bar{C}^{\alpha} \,
  \chi_{\alpha , i} (A, \mathcal{B}) \,
R^i_{\beta}(A+\mathcal{B}) C^{\beta},
\\
\n{Sgf}
S_{gf} (\phi, \mathcal{B})
&=
B^{\alpha}\chi_{\alpha}(A, \mathcal{B}).
\end{align}
In the last expression $\,\chi_{\alpha} (A, \mathcal{B})\,$ are functions
lifting the degeneracy for the  action $S_0(A+\mathcal{B})$. The standard
background field gauge condition in the BFM is linear in the quantum fields
\beq
\n{GC}
 \chi_\al(A,\mathcal{B}) \,=\,F_{\al i}(\mathcal{B})A^i\,.
\eeq

 The action \eqref{FP}
is invariant under the BRST symmetry \cite{BRS,T}
\beq
\n{BRST}
\de_{B} \phi^A = {s}^{A}(\phi, \mathcal{B}) \mu,
\qquad
\vp\big({s}^{A}(\phi, \mathcal{B}) \big) = \vp_{A} +1\,,
\eeq
where
\begin{align}
\n{si}
s^{A}(\phi, \mathcal{B})
 = & \Big(
R^i_{\alpha}(A+\mathcal{B}) C^{\alpha},
\, 0, \,-\frac{1}{2} F^{\al}_{\be\ga} C^\ga C^\be (-1)^{\vp_\be }, 
(-1)^{\vp_\al} B^\al
\Big)
\end{align}
and $\mu$ is a constant Grassmann parameter with $\,\vp(\mu) = 1$.
One can write \eqref{si} as generator of BRST transformations,
\beq
\hat{s}(\phi, \mathcal{B} ) =
 \frac{\overleftarrow{\de}}{\de \phi^{A}}
s^{A}(\phi, \mathcal{B} )
.
\label{shat}
\eeq
Then, the action \eqref{FP} can be written in the form
\beq
\n{FP+GF}
S_{FP} (\phi, \mathcal{B})
\,=\, S_{0} (A + \mathcal{B}) + \Psi(\phi, \mathcal{B} )
\, \hat{s}(\phi, \mathcal{B} )\,,
\eeq
where
\beq
\n{GFF}
\Psi(\phi, \mathcal{B} ) =
\bar{C}^{\alpha} \chi_{\alpha} (A, \mathcal{B}),
\eeq
is the gauge fixing functional. The transformation (\ref{BRST})
is nilpotent, that means $\hat{s}^2 = 0$. Taking into
account that
$S_{0} (A + \mathcal{B}) \,\hat{s}(\phi, \mathcal{B} ) = 0$,
the BRST symmetry of $S_{FP}(\phi, \mathcal{B})$ follows
immediately
\beq
\n{BRSTSFP}
S_{FP} (\phi, \mathcal{B}) \,\hat{s}(\phi, \mathcal{B} ) = 0
\, .
\eeq

Due to the presence of external vector field $ \mathcal{B}^i$, the
Faddeev-Popov action obeys an additional local symmetry known
as the background field symmetry,
\beq
\n{InSFP}
\de_{\omega}  S_{FP}(\phi, \mathcal{B}) =0,
\eeq
which is related to the background field transformations
\beq
\n{TT}
&&\de^{(c)}_{\omega} \mathcal{B}^i=
R^i_{\alpha}(\mathcal{B})\omega^{\alpha},
\nn
\\
&&\de_\om^{(q)} A^i
= \left[ R^i_\al (A+\mathcal{B})
- R^i_\al (\mathcal{B}) \right] \om^\al ,
\nn
\\
&&\de_\om^{(q)} B^\al
= -F^{\al}_{\ga\be}  B^\be \om^\ga,
\\
&&\de_\om^{(q)} C^\al = -F^{\al}_{\ga\be}  C^\be \om^\ga
 (-1)^{\vp_{\ga}},
\nn
\\
&&\de_\om^{(q)} \bar{C}^\al
= -F^{\al}_{\ga\be}  \bar{C}^\be \om^\ga  (-1)^{\vp_{\ga}}.
\nn
\eeq
Here the subscript $(c)$ is used to indicate the background field
transformations in the sector of external (classical) fields while
the $(q)$ in the sector of quantum fields (integration variables in
functional integral for generating functional of Green functions).
The symbol $\de_{\omega}$ means  the combined background
field  transformations $\,\de_\om  = \de_\om^{(c)} +\de_\om^{(q)} $.
Note that in deriving (\ref{InSFP}) the transformation rule for the
gauge fixing functions 
(\ref{GC})
\beq
\n{GFT}
\de_\om \chi_\al(\phi,\mathcal{B})
\,=\,  - \chi_\be (\phi,\mathcal{B})  F^{\be}_{\al\ga} \om^\ga\, ,
\eeq
under the background field transformations (\ref{TT}) is assumed.
It is useful to introduce the generator of the background field
transformation $\,{\hat{\cal R}}_{\omega}(\phi, \mathcal{B})$,
\begin{align}
{\hat{\cal R}}_{\omega}(\phi, \mathcal{B})
&=
\int dx \;\Big( \frac{\overleftarrow{\de}}{\de  \mathcal{B}^a_{\mu}}
\de^{(c)}_{\omega} \mathcal{B}^a_{\mu}
+ \frac{\overleftarrow{\de}}{\de \phi^{i}}\de_{\omega}^{(q)}\phi^i\Big)
\nn \\ &
= 
{\hat{\cal R}}^{(c)}_\om( \mathcal{B})+{\hat{\cal R}}^{(q)}_\om(\phi),
\label{Rcq}
\end{align}
where $ \phi^j{\hat{\cal R}}^{(q)}_{\omega}(\phi) = {\hat{\cal R}}^j_{\omega}(\phi) $
and
\beq
{\hat{\cal R}}^{j}_{\omega}(\phi)
=
\left(
{\cal R}^{(q)}_{\omega}(A)
,\, {\cal R}^{(q)}_{\omega}(B)
,\,{{\cal R}}^{(q)}_{\omega}(C)
,\,{{\cal R}}^{(q)}_{\omega}({\bar C}) \right).
\label{Rs}
\eeq
Using the new notations (\ref{Rcq}), the background field invariance of the
Faddeev-Popov action (\ref{InSFP}) rewrites as
\beq
\n{BFSSFP}
 S_{FP}(\phi, \mathcal{B}) \,{\hat{\cal R}}_{\om}(\phi, \mathcal{B})=0.
\eeq

The symmetries (\ref{BRSTSFP}) and (\ref{BFSSFP}) of the
Faddeev-Popov  action  lead to the two very important properties
at the quantum level. In order to reveal these consequences we
have to introduce the extended generating functional of Green functions
in the background field method in the form of functional integral
\begin{align}
\n{GFGF}
Z(J,\phi^*, \mathcal{B}) =
\int \mathcal{D} \phi  \,
\text{exp} \left\{\frac{i}{\hbar}
\left[ S_{FP} (\phi, \mathcal{B})
+ \phi^*(\phi {\hat s})+J \phi\right] \right\}
= \text{exp}\left\{\frac{i}{\hbar}
W (J, \phi^*,\mathcal{B}) \right\}\,,
\end{align}
where $W = W(J, \phi^*, \mathcal{B})$ is the extended generating functional of connected
Green functions and
\beq
J_{A}
= \left(J_i,\, J^{B}_{\al}, \,\bar{J}_{\al},\, J_{\al} \right)
\eeq
are the external sources to the
fields $\,\phi^A$ $\left( \vp(J_{A}) = \vp_{A} \right)$.
Furthermore, the new quantities (antifields)
$\phi^*_A$, with $\vp(\phi^*_A) = \vp_{A}+1$, are the sources
of the BRST transformations.

The introduction of antifields enable one to simplify the use of the BRST
symmetry at the quantum level. The next step is to introduce the extended
effective action $\Ga=\Ga(\Phi,\phi^*, \mathcal{B})$ through the Legendre
transformation of $W(J, \phi^*,\mathcal{B})$
\beq
\Ga(\Phi,\phi^*, \mathcal{B})
=W(J, \phi^*,\mathcal{B})-J\Phi,
\label{EA}
\eeq
where
\beq
\Phi^A = \frac{\delta_l W}{\delta J_A}
\quad
\mbox{and}
\quad
\frac{\delta_r \Gamma}{\delta \Phi^A} =-J_A.
\label{EA-ids}
\eeq
From one hand, one can prove that the BRST symmetry~(\ref{BRSTSFP}) of $S_{FP}$
results in the Slavnov-Taylor identity \cite{Taylor,Slavnov}
\beq
\n{STG}
\frac{\delta_r \Gamma}{\delta \Phi^A} \frac{\delta_l \Gamma}{\delta \phi^*_ A}
=0.
\eeq
On the other hand, the background field symmetry (\ref{BFSSFP})
of  $S_{FP}$  leads to the symmetry of the effective action under
the background field transformations,
\beq
\n{BSG}
\tilde{\Ga}(\Phi, \mathcal{B})
{\hat{\cal R}}_{\om}(\Phi, \mathcal{B})=0,
\quad
\tilde{\Ga}(\Phi, \mathcal{B}) =\Ga(\Phi,\phi^*=0, \mathcal{B}).
\eeq

The fundamental object of the background field method is the background
effective action $\Ga( \mathcal{B}) \equiv \tilde{\Ga}(\Phi=0, \mathcal{B})$.
Thanks to the linearity of $ {\hat{\cal R}}_{\om}(\Phi, \mathcal{B})$ with
respect to the mean fields $\Phi^ i$, from (\ref{BSG}) it follows
\beq
\n{BSbG}
\de^ {(c)}_{\omega}  \Gamma(\mathcal{B}) =0,
\qquad
\de^{(c)}_{\omega} \mathcal{B}^i=
R^i_{\alpha}(\mathcal{B})\omega^{\alpha},
\eeq
i.e. the background effective action is a gauge invariant functional of the
external field $\mathcal{B}^i$.

The last important feature of the Faddeev-Popov quantization is related to the universality of the $S$-matrix,
that is independent on the choice of the gauge fixing. According to
the well-known result \cite{KT}, the universality of the $S$-matrix
is equivalent to the gauge fixing independent vacuum functional.
In the background field formalism this functional
is defined starting from (\ref{GFGF}) as
\begin{align}
{Z}_{\Psi} (\mathcal{B}) = Z(\mathcal{B}, J=\phi^*=0)
= \int {\cal D} \phi\,\exp
\left\{\frac{i}{\hbar}\,S_{FP}(\phi, \mathcal{B})\right\}.
\label{VacFun}
\end{align}
Regardless this object depends on the background field,
it is constructed for a certain choice of gauge $\Psi (\phi, \mathcal{B})$.
However, it can be shown to be independent on this choice.
Without the presence of background field, the discussion
of this issue in usual QFT and in the FRG approach can
be found in Ref.~\cite{LSh}. Here we generalize it for the
\BFM case.

Taking an infinitesimal change of the gauge fixing functional,
$\,\Psi(\phi,\mathcal{B})\rightarrow \Psi(\phi,\mathcal{B})
+\delta\Psi(\phi,\mathcal{B})$, we get
\begin{align}
\label{WIZvd}
{Z}_{\Psi+\delta\Psi} (\mathcal{B}) = &
\int {\cal D}\phi\, 
\exp \Big\{ \frac{i}{\hbar}\,\Big[
S_{FP}({\phi}, \mathcal{B})
+ \de \Psi ({\phi}, \mathcal{B}) \hat{s} ({\phi}, \mathcal{B})
\Big]\Big\}
.
\end{align}
Then, after a change of variables in the form of BRST transformation
\eq{BRST} but with replacement of the constant parameter $\mu$ by the
functional
\beq
\mu ({\phi}, \mathcal{B}) = \frac{i}{\hbar}\, \de \Psi ({\phi}, \mathcal{B})
,
\eeq
one can show that
\beq
{ Z}_{\Psi + \delta\Psi} (\mathcal{B}) = { Z}_{\Psi} (\mathcal{B}),
\label{Inva-Vac}
\eeq
which is the starting point for the proof of the gauge fixing
independence of the $S$-matrix \cite{KT,Tyutin}. In the next
sections we shall see how this and other features for the case of the Yang-Mills theory look in the framework of the FRG approach.

\section{Background average effective action }
\label{Sec3}

In this section we shall discuss the use of the BFM
 applied to the FRG, following the original publication on this
subject by Reuter and Wetterich \cite{Reuter} for the case of pure Yang-Mills theory
with the action
\beq
\label{ActionYM}
S_{0} (A) = - \frac{1}{4} F^a_{\mu\nu}(A) F^a_{\mu\nu} (A),
\eeq
where
$\,F^a_{\mu\nu}(A) = \pa_\mu A^a_\nu - \pa_\nu A^a_\mu
+ gf^{abc} A^b_\mu A^c_\nu\,$
is the field strength for the non-Abelian  vector field $A_\mu$ and $g$ is coupling
constant.
The correspondence
with the notations used in Sec.~2 reads
\begin{align}
&
A^i \rightarrow A^a_\mu ,
\qquad  
\mathcal{B}^i \rightarrow \mathcal{B}^a_\mu ,
\qquad
F^{\al}_{\be\ga} \rightarrow f^{abc} ,
\nn \\ & 
 R^i_\al(A) \rightarrow D^{a b}_\mu(A)
= \de^{ab} \partial_\mu + gf^{acb} A^c_\mu.
\end{align}
Here the structure coefficients $f^{abc}$ of the gauge group are
constant. The action~\eqref{ActionYM} is invariant under the gauge
transformations defined by the generator $D^{ab}_\mu (A)$ with an
arbitrary gauge function $\om^a$ with $\vp(\om^a) =0$. In the
Faddeev-Popov quantization, the Grassmann parity of the fields
$\,\phi^A = ( A^a_\mu, B^a, C^a, \bar{C}^a )$ is, respectively,
$\vp_A = (0,0,1,1)$.

The background field formalism for Yang-Mills theory comprises
the definition of the background field transformation
\beq
\label{YM-back}
&&
\de_\om^{(c)} \mathcal{B}^a_\mu = D^{ab}_\mu (\mathcal{B}) \om^b ,
\qquad
\de_\om^{(q)} A^a_\mu \,=\,g f^{abc} A^b_\mu \om^c ,
\nonumber
\\
&&
\de_\om^{(q)} B^a = gf^{abc}  B^b \om^c ,
\qquad
\de_\om^{(q)} C^a = gf^{abc} C^b \om^c ,
\qquad
\nn
\\
&&
\de_\om^{(q)} \bar{C}^a = gf^{abc} \bar{C}^b \om^c  .
\mbox{\qquad}
\eeq
Note that the generator of the
 transformation in the sector of fields $A^a_\mu$ reads
\beq
D^{ab}_\mu (A + \mathcal{B}) - D^{ab}_\mu (\mathcal{B})
= gf^{acb} A^c_\mu ,
\eeq
and thus all the quantum fields transform according the same rule.
The standard choice of the gauge-fixing function is
\beq
\chi^a (A,\mathcal{B}) = D^{ab}_\mu (\mathcal{B}) A^b_{\mu}.
\eeq
It leads to the tensor transformation rule for $\chi^a (A,\mathcal{B})$ under
the background field transformation,
\beq
\delta_{\omega}\chi^a (A,\mathcal{B})=gf^{abc} \chi^b (A,\mathcal{B}) \om^c  .
\eeq

The main point
of the FRG approach is the introduction of the scale-dependent
regulator action $S_k (\phi,\mathcal{B})$, in the framework of
the background field method. Let us choose the regulator action
for the quantum fields $\,A^a_\mu\,$ and $\,C^a,\,\bar{C}^a$ in the form
\begin{align}
\n{SkB}
S_k (\phi,\mathcal{B})
= &
\frac12 A^{a}_{\mu}\,
R^{(1)\; ab}_{k \;\mu \nu}(D_T(\mathcal{B})) \,A^{b}_{\nu}
+
{\bar C}^a\,
R^{(2)\; ab}_{k}(D_S(\mathcal{B})) \,C^b
.
\end{align}
The regulator functions depend on the external field through the covariant derivatives of tensor $D_T$ and scalar  $D_S$
fields
\begin{align}
\n{AR}
({ D}_T {\cal (B)})_{\mu\nu}^{ab}
=& - \,\eta_{\mu\nu} (D^2)^{ab}
+ 2 g f^{acb} F^c_{\mu\nu} ({\cal B}),
\qquad 
  (D^2)^{ab} =  D^{ac}_\rho ({\cal B}) D^{cb}_\rho ({\cal B}),
 \\
({ D}_S {\cal (B)})^{ab}
= & - \,(D^2)^{ab}  .
\end{align}
The form of these functions can be chosen e.g. as in \cite{Reuter},
\beq
\n{R}
R_k(z) = Z_k \frac{ze^{-z/k^2}}{1-e^{-z/k^2}},
\eeq
with $Z_k$ corresponding to the wave function renormalization.

Let us consider the variation of the regulator action \eq{SkB} under
the background field transformations \eqref{TT} in the first order
approximation, $R_k(z)=Z_k z$.  The first term in \eq{AR} can be rewritten
through integration by parts, as follows
\beq
- \,A^a_\mu  \eta_{\mu\nu} (D^2)^{ab}
A^b_\nu
=
\chi^c_{\rho\mu} (A, {\cal B })\, \chi^c_{\rho\mu} (A, {\cal B }),
\eeq
where
\beq
\chi^a_{\rho\mu} (A, {\cal B })
\equiv D^{ab}_\rho {\cal (B) } A^b_\mu.
\eeq
The transformation rule for $\,\chi^a_{\rho\mu} (A, {\cal B })\,$ under
the background field transformation is very close to \eq{GFT}. It has the form
\beq
\de_\om \, \chi^a_{\rho\mu} (A, {\cal B }) =
g f^{acb} \chi^c_{\rho\mu} (A, {\cal B }) \om^b.
\eeq
As consequence, we find the first term invariance
\begin{align}
\de_\om ( - A^a_\mu  \eta_{\mu\nu} (D^2)^{ab} A^b_\nu )
&=
\de_\om (\chi^a_{\rho\mu} (A, {\cal B }) \chi^a_{\rho\mu} (A, {\cal B }))
\nn \\&=
2 g f^{acb} \chi^a_{\rho\mu}
(A, {\cal B }) \chi^c_{\rho\mu} (A, {\cal B }) \om^b 
\nn \\&
= 0\,.
\end{align}
Furthermore, taking into account that
\beq
\de^{(c)}_\om F^a_{\mu\nu} ({\cal B})
= g f^{acb} F^c_{\mu\nu} ({\cal B}) \om^b,
\eeq
for the second term in \eq{AR}, we have
\begin{align}
 \de_\om \, \big(f^{acb} A^a_\mu F^c_{\mu\nu} ({\cal B}) A^b_\nu\big)
 = &
 gA^a_\mu A^b_\nu F^c_{\mu\nu}
\big(f^{ace} f^{ebd} +
+ f^{abe} f^{edc} 
+ f^{ade} f^{ecb} \big) = 0,
\nn
\end{align}
because of the Jacobi identity. The invariance holds also
for the ghost regulator, as one can easily verify. In this approximation the scale-dependent
action $S_k (\phi,\mathcal{B})$ obeys the background field symmetry,
$\delta_{\omega}S_k (\phi,\mathcal{B})=0$.

The same consideration can be done for the terms of the higher
orders in $z$. Thus, we can ensure that the invariance is maintained
in all orders. With these results the action (\ref{SkB}) is invariant
under the background field transformations,
\beq
\n{CISK}
\de_\omega  S_k (\phi,\mathcal{B}) = 0.
\eeq

The full action $S_{k \;FP} =S_{k \;FP} (\phi, \mathcal{B}) $
is constructed by the rule
\beq
\n{FPSK}
S_{k \;FP} (\phi, \mathcal{B})
= S_{FP} (\phi, \mathcal{B}) + S_k (\phi,\mathcal{B})\,,
\eeq
where $S_{FP} (\phi, \mathcal{B})$ is the Faddeev-Popov
action (\ref{FP}). Using the action (\ref{FPSK}), the generating
functional of Green function is given by the following functional
integral:\footnote{From here we adopt units in which $\hbar=1$.}
\beq
\n{dZk}
Z_k(J,\mathcal{B})
 = 
\int {\cal D} \phi \,
\exp\Big\{i[ S_{FP} (\phi, \mathcal{B})
+ S_k (\phi,\mathcal{B})
+J\phi]\Big\}
 = \exp\big\{iW_k(J,\mathcal{B})\big\},
\eeq
where $W_k=W_k(J,\mathcal{B})$ is the generating functional
of connected Green functions. The main object of the FRG
approach in the background field method is the \bAEA
$\Gamma_k=\Gamma_k(\Phi, \mathcal{B} )$, defined through
the Legendre transform of $\,W_k$,
\beq
&&
\Gamma_k(\Phi, \mathcal{B} )
\,=\,W_k(J,\mathcal{B})-J\Phi,
\eeq
where
\beq
\Phi^A = \frac{\delta_l W_k}{\delta J_A}
\nn
\eeq
and
\beq
\frac{\delta_r \Gamma_k}{\delta \Phi^A}=-J_A.
\nn
\eeq

The effective average action can be presented as a sum of
the regulator action of the mean field and the quantum correction,
\beq
&&
\Gamma_k(\Phi, \mathcal{B} )
= S_k(\Phi, \mathcal{B})
 + {\bar \Gamma}_k(\Phi, \mathcal{B} ).
\label{EA-ca}
\eeq
The functional $\bar{\Ga}_k$ satisfies the flow equation, or
the Wetterich  equation \cite{Wetterich,Reuter},
\beq
\n{FEG}
\pa_t {\bar \Gamma}_k(\Phi, \mathcal{B} )
=
\frac{i}{2}\sTr \Bigg\{
\frac{\pa_t R_{k}( \mathcal{B})}
{\,\,
{\bar \Gamma}^{''}_k(\Phi, \mathcal{B})
\,+\,R_k( \mathcal{B})\,}\Bigg\}.
\eeq
In (\ref{FEG}) $ \pa_t =k\frac{d}{d k} $ and the symbol $\sTr $ means the
functional supertrace, this last is necessary due to the presence
of quantum fields $A^a_\mu\,$ and $\,C^a,\,\bar{C}^a$, with different
Grassmann parity. Another important notation is
\beq
\Big({\bar \Gamma}^{''}_k(\Phi, \mathcal{B})\Big)_{AB}
=
\frac{\delta_l}{\delta\Phi^A}
\left(\frac{\de_r {\bar\Ga}_k(\Phi, \mathcal{B} )}{\de\Phi^B}\right)
\eeq
for the matrix of the second order functional derivatives with
respect to the mean fields $\Phi$.

As we have seen above, because of the invariance of the
scale-dependent regulator term (\ref{SkB}), the full action
(\ref{FPSK}) is invariant under the background field
transformations (\ref{InSFP}),
\begin{align}
\n{nISk}
\delta_{\omega}S_{k\;FP}(\phi,\mathcal{B})
=\delta_{\omega}S_k(\phi,\mathcal{B})
=
S_k(\phi,\mathcal{B}) {\hat{\cal R}}_{\omega}(\phi, \mathcal{B})=0.
\end{align}
At the quantum level (\ref{nISk}) provides the invariance of the
background
average effective action $\Gamma_k(\Phi, \mathcal{B} )$.
Indeed, variation of $Z_k(J,\mathcal{B})$ with respect to the
external field $\mathcal{B}^a_\mu$ reads
\beq
\n{vZk}
\delta^{(c)}_{\omega}Z_k(J,\mathcal{B})=iJ_A
\;\!{\cal R}^A_{\omega}\left(\frac{\delta_l Z_k}{i\delta J}\right).
\eeq
In terms of the functional $W_k(J,\mathcal{B})$ the relation
(\ref{vZk}) rewrites
\beq
\n{vWk}
\delta^{(c)}_{\omega}W_k(J,\mathcal{B})=
J_A\;\!{\cal R}^A_{\omega}\left(\frac{\delta_l W_k}{\delta J}\right).
\eeq
As a consequence of (\ref{vWk}), the background average
effective action is invariant under the background field
transformations,
\beq
\n{vGk}
&&\delta_{\omega}\Gamma_k(\Phi,\mathcal{B})=0.
\eeq
In terms of the functional $ {\bar \Gamma}_k(\Phi, \mathcal{B} )$
the relation (\ref{vGk}) becomes
\beq
\n{vGka}
\delta_{\omega}{\bar \Gamma}_k(\Phi, \mathcal{B} )=0.
\eeq
Thus, the background field symmetry is preserved for the background
average effective action $\,{\bar\Ga}_k(\Phi, \mathcal{B} )$,
confirming the main statement of  the  paper \cite{Reuter}.

For the functional
${\bar \Ga}_k( \mathcal{B})={\bar \Ga}_k(\Phi=0, \mathcal{B})$,
the background field symmetry is preserved as well due to linearity
of the background field symmetry
\beq
\n{vGkab}
\delta^ {(c)}_{\omega}{\bar \Gamma}_k( \mathcal{B})=0,
\eeq
in agreement  with (\ref{BSbG}).  In particular this means that the
flow equation for~$\,{\bar \Gamma}_k( \mathcal{B})$,
\beq
\pa_t {\bar \Gamma}_k(\mathcal{B} )
=
\frac{i}{2}\sTr \Bigg\{
\frac{\pa_t R_{k}( \mathcal{B})}
{\,\,
{\bar \Gamma}^{''}_k(\Phi,\,\mathcal{B})\big|_{\Phi =0}
\,+\,R_k( \mathcal{B})\,}\Bigg\},
\eeq
maintains the  background field symmetry.

\section{Background invariant regulator functions}
\label{Sec4}

The prove of invariance of $S_k$ under background field
transformations (\ref{CISK}) is based on the certain form
of the regulator functions and its arguments. In particular,
the regulator functions (\ref{R}) with argument (\ref{AR}) by itself are
not invariant under background field transformations
$\,\de^{(c)}_{\om} R^{(1)\; ab}_{k \;\mu \nu}(D_T(\mathcal{B}))\neq 0$,
$\de^{(c)}_{\om}R^{(2)\; ab}_{k}(D_S(\mathcal{B}))\neq 0$.
In this section we shall discuss the background field symmetry
of the background average effective action and formulate a possible
restriction on the regulator functions in the scale-dependent action
$S_k$ in the general settings that allow us to arrive at the
invariance of the \bAEA under background field transformations.

Consider the scale-dependent regulator action
$S_k = S_k (\phi, \mathcal{B})$ in the background field
formalism, including the ghost sector,
\begin{align}
S_k (\phi, \mathcal{B}) =
\frac{1}{2}  A^{a}_{\mu}
R^{(1)\; ab}_{k \;\mu \nu} (\mathcal{B})
A^{b}_{\nu}
\,+\,
\bar{C}^{a}
R^{(2)\; ab}_{k}(\mathcal{B})
C^{b},
\end{align}
where  $R^{(1)\; ab}_{k \;\mu \nu}(\mathcal{B})\,$ and
$R^{(2)\; ab}_{k}(\mathcal{B})\,$  are the regulator functions.
We assume that they are local functions of external fields
$\mathcal{B}^a_{\mu}$ and their partial derivatives.
The full action has a standard FRG form
\beq
\label{fullA}
S_{kFP} (\phi, \mathcal{B})
=
S_{FP} (\phi, \mathcal{B})+S_k (\phi, \mathcal{B}).
\eeq
Due to the background field symmetry of the Faddeev-Popov
action (\ref{InSFP}), the full action (\ref{fullA}) will be invariant
under the background field transformations (\ref{TT}), if  the
scale-dependent regulator action $ S_k = S_k (\phi, \mathcal{B})$
satisfies the equation
\beq
\n{ISK}
\de_{\omega} S_k (\phi, \mathcal{B}) = 0.
\eeq
Using the explicit form of the background field transformations
(\ref{TT}) the variation of $ S_k (\phi, \mathcal{B})$ reads
\begin{equation}
\begin{split}
\de_{\omega} S_k (\phi, \mathcal{B})
= &
 \frac{1}{2} A^ a_{\mu}
\Big[g
\Big(f^ {adc}\om^ d R_{k\;\mu\nu}^{(1)\; cb}(\mathcal{B})
- R_{k\; \mu\nu}^ {(1)\; ac} (\mathcal{B}) f^ {cdb} \om^d
\Big)
+\delta^{(c)}_{\omega} R^{(1)\; ab}_{k \;\mu \nu} (\mathcal{B})
\Big] A^b_{\nu}
\\ &
\n{ISKa}
+
\int dx \, {\bar C}^a\Big[g
\Big(f^{adc} \omega^d
R^{(2)\; cb}_{k}(\mathcal{B})
- R^{(2)\;ac}_{k}(\mathcal{B}) f^{cdb} \omega^d
\Big)
+\de^{(c)}_{\omega} R^{(2)\;ab}_{k}(\mathcal{B})\Big]
C^b.
\end{split}
\end{equation}
From Eq.~(\ref{ISKa}) follows that (\ref{ISK}) is satisfied if
\begin{align}
\n{R1}
&
g \Big( f^ {adc}\omega^d R_{k\;\mu\nu}^ {(1)\;cb}(\mathcal{B})
- R_{k\;\mu\nu}^{(1)\; ac}(\mathcal{B}) f^ {cdb} \omega^ d
\Big) + \delta^{(c)}_{\omega}R_{k\; \mu\nu}^ {(1)\;ab}(\mathcal{B})
=0,
\\
\n{R2}
&
g\left(f^{adc} \omega^d
R^{(2)\;cb}_{k}(\mathcal{B})
- R^{(2)\;ac}_{k }(\mathcal{B})
f^{cdb} \omega^d\right)
+\de^{(c)}_{\omega} R^{(2)\;ab}_{k}(\mathcal{B})
=0.
\end{align}
Any solution of these equations provides the invariance of $S_k$
under background field transformations. Let us consider the case
when regulator functions are invariant under background
transformations of external field $\mathcal{B}^a_\mu$,
\beq
\n{Rin}
\delta^{(c)}_{\omega}R_{k\; \mu\nu}^ {(1)\; ab}(\mathcal{B})
=0,
\qquad
\de^{(c)}_{\omega} R^{(2)\;ab}_{k}(\mathcal{B})
=0.
\eeq
Due to the arbitrariness in the choice of the functions
$\omega^{a}(x)$, from (\ref{R1}), (\ref{R2}) and
(\ref{Rin}) follow the relations
\beq
\big[t^d,R_{k}^ {(1)\;\mu\nu}(\mathcal{B} ) \big]_{ab}=0,
\qquad
\big[t^d,R_{k}^ {(2)\;\mu\nu}(\mathcal{B}) \big]_{ab}=0,
\eeq
for the generators $(t^a)_{bc} = f^{bac}$ of the Lie group.
Therefore, we see that the regulator functions commute with all the
generators of Lie group. Then, applying the Shur's lemma we find
\begin{align}
R_{k\;\mu\nu}^{(1)\;ab}(\mathcal{B}) =&
\delta^{ab} R_{k \;\mu\nu}^ {(1) }(D(\mathcal{B})),
\nn \\ 
R_{k}^{(2)\;ab}(\mathcal{B})
=& \delta^{ab} R_{k}^ {(2)}(D(\mathcal{B})),
\end{align}
where the quantities $R_{k\; \mu\nu}^ {(1)}(D(\mathcal{B}))$
and $R_{k}^{(2)}(D(\mathcal{B}) )$ are scalars with respect
to the background transformations of external field $\mathcal{B}^a_\mu$.
It means that the arguments of these quantities should be scalars
as well.  It is easy to construct an example of such kind of
a scalar argument, $D(\mathcal{B})=F^{a}_{\mu\nu}(\mathcal{B})
D^{ab}_{\mu}(\mathcal{B})\mathcal{B}^b_{\nu}$, where
$F^{a}_{\mu\nu}$ is defined in (\ref{ActionYM}).

So, in the case under consideration, the scale-dependent regulator
action has the form
\begin{align}
S_k (\phi,\mathcal{B})
= &
 \frac{1}{2}  A^{a}_{\mu}
R^{(1)}_{k \; \mu \nu}(D(\mathcal{B}) )
A^{a}_{\nu}
+ \bar{C}^{a}(x) R^{(2)}_{k}(D(\mathcal{B}) )
C^{a},
\end{align}
maintaining the background field symmetry $\de_\om S_k (\phi,\mathcal{B}) =0$.

\section{Gauge dependence of \bAEA}
\label{Sec5}

Here the problem of gauge dependence of \bAEA will be discussed in general
setting of Sec. 2
The regulator action $S_k$ is invariant under the background
transformations (\ref{CISK}), but not under the
BRST transformations,
\beq
S_k (\phi,\mathcal{B}) {\hat s}(\phi,\mathcal{B}) \neq 0.
\eeq
Let us  discuss the implications of this fact for the
gauge dependence problem of the background average effective action.
Consider the extended generating functional of Green functions $Z_k(J,\phi^*, \mathcal{B})$, and the extended generating
functional of connected Green functions $\,W_k(J, \phi^*, \mathcal{B})$,
\begin{align}
\label{ZkWk}
Z_k(J,\phi^*, \mathcal{B})
= &
\int { \cal D} \phi\exp\{i[S_{FP}(\phi, \mathcal{B})
+ S_k(\phi, \mathcal{B})
+ J\phi
+\phi^*({\hat s}\phi)]\}
=
\exp\{iW_k(J,\phi^*, \mathcal{B})\},
\end{align}
As the first step we derive the modified Ward identity for the FRG
in the BFM which is a consequence
of the BRST invariance of the action $S_{FP}(\phi, \mathcal{B})$ (\ref{BRSTSFP}).
Making use the change of variables in the form of the BRST transformations
in the functional integral (\ref{ZkWk}),
$\phi^A\rightarrow \varphi^A(\phi)=\phi^A+({\hat s}\phi^A)\mu$, and taking into account
the triviality of the corresponding Jacobian if the conditions
\beq
\label{nr}
(-1)^{\varepsilon_i}\frac{\delta_l R^i_{\alpha}}{\delta A^i}+
(-1)^{(\varepsilon_{\alpha}+1)}F^{\beta}_{\beta\alpha}=0
\eeq
are satisfied (for detailed discussion of this point see \cite{GLSh}), we arrive at
the relation
\begin{align}
\nn
0 = &\int D\phi \;\big(J_A+S_{k,A}(\phi, \mathcal{B})\big)({\hat s}\phi^A)
\exp\{i[S_{FP}(\phi, \mathcal{B})
+ S_k(\phi, \mathcal{B})
+ J\phi+\phi^*({\hat s}\phi)]\}
\nn \\ 
\label{rel}
= &\left[J_A+S_{k,A}\left(\frac{\delta_l}{i\delta J}, \mathcal{B}\right) \right]
\int D\phi ({\hat s}\phi^A)\exp\{i[S_{FP}(\phi, \mathcal{B})
+ S_k(\phi, \mathcal{B})+ J\phi+\phi^*({\hat s}\phi)]\}.
\end{align}
From (\ref{rel}) it follows the modified Ward identity for the extended
generating functional of Green functions $Z_k(J,\phi^*,\mathcal{B})$
\beq
\label{WIZk}
\left[J_A+S_{k,A}\left(\frac{\delta_l}{i\delta J}, \mathcal{B}\right) \right]
\frac{\delta_l}{\delta\phi^*_A}Z_k(J,\phi^*, \mathcal{B})=0.
\eeq
This identity in terms of the extended generating
functional of connected Green functions
$\,W_k(J, \phi^*, \mathcal{B})$ reads
\begin{align}
\label{WIWk}
\left[J_A+S_{k,A}\left(\frac{\delta_l W_k}{\delta J}+
\frac{\delta_l}{i\delta J},\mathcal{B}\right) \right]
\frac{\delta_l}{\delta\phi^*_A}W_k(J,\phi^*, \mathcal{B})=0.
\end{align}
Introducing the generating functional of vertex functions
$\Gamma_k=\Gamma_k(\Phi,\phi^*, \mathcal{B})$ with the help
of Legendre transformation of $W_k=W_k(J,\phi^*, \mathcal{B})$
\begin{align}
&\Gamma_k(\Phi,\phi^*, \mathcal{B})=W_k(J,\phi^*, \mathcal{B})-J_A\Phi^A, 
\qquad
\Phi^A=
\frac{\delta_l W_k}{\delta J_A}, 
 \\ &
\frac{\delta_r \Gamma_k}{\delta \Phi^A}=-J_A,\quad
\frac{\delta_l \Gamma_k}{\delta \phi^*_A}=
\frac{\delta_l W_k}{\delta \phi^*_A} \nn,
\end{align}
the modified Ward identity rewrites in the form
\beq
\label{WIGk}
\frac{\delta_r \Gamma_k}{\delta \Phi^A}
\frac{\delta_l \Gamma_k}{\delta \phi^*_A}=
S_{k,A}({\hat\Phi}, \mathcal{B})\frac{\delta_l \Gamma_k}{\delta \phi^*_A},
\eeq
where the notation
\beq
\label{hatPhi}
{\hat\Phi}^A=\Phi^A+i\,(\Gamma_k^{''-1})^{AB}\,
\frac{\delta_l}{\delta\Phi^B},
\eeq
has been used. The matrix \ $(\Gamma_k^{''-1})$ \ is inverse to the
matrix $\Gamma_k^{''}$, the last has elements
\begin{align}
&
(\Gamma_k^{''})_{AB} =
\frac{\de_l}{\de\Phi^A}
\Big(\frac{\de_r\Gamma_k}{\de\Phi^B}\Big)
\,,\quad \mbox{i.e.,} \nn \\&
\big(\Gamma_k^{''-1}\big)^{AC}\cdot
\big(\Gamma_k^{''}\big)_{CB}\,=\,\de^A_{\,B}\,.
\end{align}

Now consider the variation of the extended generating functional of Green
functions under infinitesimal
variation of the gauge fixing functional,
$\,\Psi(\phi,\mathcal{B})\rightarrow \Psi(\phi,\mathcal{B})
+\delta\Psi(\phi,\mathcal{B})$. We find
\begin{align}
\label{varZk}
\delta Z_k(J, \phi^*,\mathcal{B})
= &  i\int D\phi \big(\delta\Psi(\phi,\mathcal{B}){\hat s}(\phi, \mathcal{B})\big)
\exp\{i[S_{FP}(\phi, \mathcal{B})
\nn \\ &
+ S_k(\phi, \mathcal{B})
+ J\phi+\phi^*({\hat s}\phi)]\}.
\end{align}
Now take into account that the functional integral of total variational
derivative is zero we have the relation
\begin{align}
0 = &  \int D\phi \frac{\delta_r}{\delta\phi^A}
\Big[ \big(\delta\Psi s^A\big)
\exp\{i[S_{FP}(\phi, \mathcal{B})
+ S_k(\phi, \mathcal{B})+
J\phi
+\phi^*({\hat s}\phi)]\}\Big]
\label{rel1}
\nn \\
= & \int D\phi\Big[i\delta\Psi s^A
\big(J_A+S_{k,A}\big)
+\big(\delta\Psi{\hat s}\big)\Big]
\exp\{i[S_{FP}(\phi, \mathcal{B})
+ S_k(\phi, \mathcal{B})
+ J\phi+\phi^*({\hat s}\phi)]\},
\end{align}
where the BRST invariance of $S_{FP}$ action, the nilpotency of BRST transformations
and the relations (\ref{nr}) have been used. From (\ref{rel1}) one has
\begin{align}
&
i\int D\phi \big(\delta\Psi(\phi,\mathcal{B}){\hat s}(\phi, \mathcal{B})\big)
\exp\{i[S_{FP}(\phi, \mathcal{B})
+ S_k(\phi, \mathcal{B})+ 
\nn \\ &
J\phi
+\phi^*({\hat s}\phi)]\}
=
\int D\phi \big(J_A+S_{k,A}\phi,\mathcal{B}\big)s^A(\phi,\mathcal{B})
\times
\nn \\ &
\times
\delta\Psi(\phi,\mathcal{B})
\exp\{i[S_{FP}(\phi, \mathcal{B})
+ S_k(\phi, \mathcal{B})
+ J\phi+\phi^*({\hat s}\phi)]\},
\end{align}
which allows to present the Eq. (\ref{varZk}) in the form closed with respect to
$Z_k(J, \phi^*,\mathcal{B})$,
\begin{align}
\label{varZk1}
\delta Z_k(J, \phi^*,\mathcal{B})
= &
- i \left[J_A+S_{k,A}\left(\frac{\delta_l}{i\delta J},\mathcal{B}
\right)\right]
\frac{\delta_l}{\delta\phi^*_A}
\delta\Psi\left(\frac{\delta_l}{i\delta J},\mathcal{B} \right)
Z_k(J,\phi^*,\mathcal{B}),
\end{align}
or, in terms of $W_k(J, \mathcal{B})$,
\begin{align}
\label{varWk1}
\delta W_k(J, \phi^*,\mathcal{B})
= &
-\, \left[J_A+S_{k,A}\left(\frac{\delta_l W_k}{\delta J}+
\frac{\delta_l}{i\delta J},\mathcal{B}\right)\right]
\,\frac{\delta_l}{\delta\phi^*_A}\,
\delta\Psi\left(\frac{\delta_l W_k}{\delta J}
+\frac{\delta_l}{i\delta J},\mathcal{B} \right).
\end{align}
In deriving (\ref{varWk1}) the modified Slavnov-Taylor identity (\ref{WIWk})
has been used.
The last equation can be rewritten for the background average effective action,
$\Gamma_k(\Phi,\phi^*,\mathcal{B})$, in the form
\begin{align}
\n{gGk}
\delta \Gamma_k(\Phi, \phi^*, \mathcal{B})
= &
\frac{\delta_r\Gamma_k}{\delta\Phi^A}
\frac{\delta_l}{\delta\phi^*_A}
\delta\Psi({\hat \Phi},\mathcal{B})
-
S_{k, A}({\hat \Phi},\mathcal{B})
\frac{\delta_l}{\delta\phi^*_A}\delta\Psi({\hat \Phi},\mathcal{B}),
\end{align}
where ${\hat \Phi}$ was introduced in (\ref{hatPhi}).
From Eq.~(\ref{gGk}) follows that
\beq
\n{gGkn}
\delta \Gamma_k(\Phi,  \phi^*,\mathcal{B})
\Big|_{\frac{\delta \Gamma_k}{\delta\Phi}=0} \neq\ 0.
\eeq
As result, the average effective action depends on gauge fixing even
on the equations of motion (on-shell) and the $S$-matrix
defined in the framework of the FRG approach is gauge dependent.

\section{Conclusions}
\label{Sec6}

We considered several aspects of \bAEA in the FRG framework.
At the first place we confirmed the well-known classical result
of \cite{Reuter} concerning the background invariance of the
regulator actions and \bAEA in the framework of the background field method for a wide class of regulator functions which include
(\ref{R}), but can be generalized to any other functions of
the arguments $z$. As a new technical result we formulated
general conditions of regulator actions being invariant with
respect to the purely background transformations.

The main motivation of this work was to check whether the
on-shell dependence of the average effective action \cite{LSh}
holds within the background field method formalism. The answer to this question
is given by the relation (\ref{gGkn}) and is strictly positive.
This output does not
contradict the recent works \cite{Wetterich-2018,CODELLO}
because in these publications the subject of study was the
gauge invariance of background average effective action,
and the question of gauge fixing dependence was not investigated.
From our viewpoint, the
on-shell gauge dependence of the average effective action
is a fundamental principal difficulty of the FRG approach
applied to the Yang-Mills theories. We have confirmed that
the situation does not improve in the background field method,
regardless of the different structure of lifting the degeneracy
of the classical action.

It is unclear whether one can achieve a reasonable physical
interpretation of the results obtained within the FRG formalism
applied to Yang-Mills theories, and therefore it makes sense
to discuss the possible ways out from this difficult situation.

Certainly the simplest way is to ignore the problem e.g. by
deciding that one special gauge fixing is ``physical'' or
``correct'', such that changing the gauge should be strictly
forbidden. As far as FRG provides valuable nonperturbative
results, the theoretically inconsistent formulation is the price
to pay for going beyond the well-defined perturbative framework.

Another possibility is to look for some observables that may
be gauge-fixing invariant. For instance, in the fixed point the
\bAEA boils down to the standard QFT effective action and
then $S$-matrix, amplitudes and all related observables are
well-defined. Unfortunately, even in the vicinity of the fixed
point this is not true due to the relation (\ref{gGkn}). Since
the search of the nonperturbative fixed point is based on the
renormalization group flows and the last are supposed to be
gauge-fixing dependent, it is unclear how the fixed-point
invariance can be actually used.

Finally, there is an alternative formulation of the FRG in gauge
theories which is gauge-fixing independent, exactly as a conventional
perturbative QFT is \cite{LSh}. This scheme is technically more
difficult, since the regulator actions are constructed in a more
complicated way, that includes composite fields. At least by now,
the disadvantage of this approach is that there is no method to
perform practical calculations.

\section*{Acknowledgements}
\noindent
P.M.L. is grateful to the Departamento de F\'{\i}sica of the Federal
University of Juiz de Fora (MG, Brazil) for warm hospitality during
his long-term visit.  The work of P.M.L. is supported partially
by the Ministry of Education and Science of
the Russian Federation, grant 3.1386.2017 and by the RFBR grant
18-02-00153. This work of I.L.Sh. was partially supported by Conselho Nacional de
Desenvolvimento Cient\'{i}fico e Tecnol\'{o}gico - CNPq under the grant 303893/2014-1 and Funda\c{c}\~{a}o de Amparo \`a Pesquisa
de Minas Gerais - FAPEMIG under the project APQ-01205-16. E.A.R. is grateful to Coordena\c{c}\~ao de Aperfei\c{c}oamento de Pessoal
de N\'{\i}vel Superior - CAPES  for supporting his Ph.D. project.




\end{document}